\documentclass[12pt,a4paper]{article}
\usepackage{amsmath}
\usepackage{amsfonts}
\usepackage{amssymb}
\usepackage{textcomp}
\usepackage[dvips]{graphicx}
\usepackage{epsfig}
\usepackage{bm}
\usepackage{dcolumn}
\usepackage{amsfonts}
\usepackage{color}
\usepackage[left=2cm,right=2cm,top=2cm,bottom=2cm]{geometry}
\usepackage[left=2cm,right=2cm,top=2cm,bottom=2cm]{geometry}
\usepackage[tableposition=top]{caption}
\usepackage{subcaption}
\DeclareCaptionLabelFormat{gostfigure}{Fig. #2}
\DeclareCaptionLabelSeparator{gost}{.~~}
\begin{document}
\title{\textbf{Anisotropic cosmological dynamics in $f(T)$ gravity in the presence of a perfect fluid}}
\author{Maria A. Skugoreva$^{1}$\footnote{masha-sk@mail.ru}, Alexey V. Toporensky$^{1,2}$\footnote{atopor@rambler.ru}\vspace*{3mm} \\
\small $^{1}$Kazan Federal University, Kremlevskaya 18, Kazan 420008, Russia\\
\small $^{2}$Sternberg Astronomical Institute, Lomonosov Moscow State University,\\
\small Moscow 119991, Russia}

\date{ \ }

\maketitle

\begin{abstract}
    We consider the cosmological evolution of a flat anisotropic Universe in $f(T)$ gravity in the presence of a perfect fluid. It is shown that the matter content of the Universe has a significant impact of the nature of a cosmological singularity in the model studied. Depending on the parameters of the $f(T)$ function and the equation of state of the perfect fluid in question the well-known Kasner regime of general relativity can be replaced by a new anisotropic solution, or by an isotropic regime, or the cosmological singularity changes its nature to a non-standard one with a finite values of Hubble parameters. Six possible scenarios of the cosmological evolution for the model studied have been found numerically.
\end{abstract}

\section{Introduction}
~~~~Recently a new class of modified gravity theories has attracted a great deal of attention. Its story goes back to the beginning of the 20th when Einstein reformulated his General Relativity (GR) in therms of torsion instead of curvature \cite{Einstein} and the Weitzenb\"{o}ck connection~\cite{Weitzenbock} instead of the Levi-Civita one. The resulting theory, known as the teleparallel equivalent of general relativity (TEGR) \cite{Pereira}, has the same equations of motion as in GR, but it has a different mathematical structure. This theory has been forgotten for a long time and only a decade ago it attracted new attention. As for TEGR itself, despite complete equivalence to GR on the level of the equations of motion, its different mathematical background leads to different global properties, such as conservation quantities, the topic which is currently the subject of intense investigations \cite{Pereira, Obukhov}. What has also become understood only after the theory's revival is that modifying TEGR the way analogous to the famous $f(R)$ modifications of GR leads to the theory (known as $f(T)$ gravity; see the review in \cite{Cai}) which is different from $f(R)$ already at the level of equations of motion \cite{FF, Ferraro1, Linder}. The function $f(T)$ is assumed to be smooth; other possible restrictions from thermodynamical considerations have been discussed in Ref.~\cite{Liu}. The equations of motion in $f(T)$ theory is of the second order, to be compared with the fourth-order equations in $f(R)$. However, $f(T)$ gravity had some still unresolved conceptional problems connected with the lack of local Lorentz covariance \cite{Lorentz1, Lorentz2} and an unknown number of dynamical degrees of freedom (see \cite{Guzman} and references therein). 

    For the equations of motion this leads to a situation when non-pathological equations require
a special (so called ``proper'' or ``good'') tetrad to be used \cite{Obukhov, Ferraro2, Tamanini, Ferraro3}, or equivalently, a non-trivial spin connection to be calculated for a given tetrad \cite{Martin}. Otherwise, the non-symmetric part of the equations of motion requires explicitly the condition $\frac{d^2f(T)}{dT^2}=0$, which means $f(T)\propto T$ and we return to unmodified TEGR. 

    Before this problem can be understood in detail, one of the possible ways to study $f(T)$ gravity is finding solutions for the cases where the ``proper'' tetrad is known. For example, it is well known that for a flat Friedmann-Lema\^{i}tre-Robertson-Walker (FLRW) cosmology with a scale factor $a$ the diagonal cartesian tetrad $(1, a, a, a)$ is a proper tetrad. This means that the general problem does not affect studies of FLRW cosmology in $f(T)$ gravity \cite{Wu, Capozziello, Setare, Feng, Biswas, Nashed, Bamba, Laur, Saridakis, Mishra, Agostino}. Moreover, it was shown that the diagonal tetrad $(1, a, b, c)$ is a proper tetrad for a flat anisotropic Universe (here $a$, $b$, $c$ are three different scale factors of the Bianchi I metric), which allows us to study corresponding anisotropic cosmological dynamics \cite{Cai, anis1, anis2, anis3, anis4, anis5, anis6, anis7, anis8, we}.    

    In our recent paper \cite{we} we have considered Bianchi I cosmological dynamics of a vacuum Universe in $f(T)=T+f_0 T^N$ gravity. We have found that, depending on the branch of solutions ($T=0$ or $T\neq 0$), the Kasner solution \cite{Kasner} is either an exact (for $T=0$ branch) or asymptotic solution (for $T\neq 0$ branch) in the high-energy regime where the corrections to Einstein gravity dominate. However, intuitively we would expect that deviations from the solutions of Einstein gravity (the Kasner solutions belong to this class) should increase as the higher order terms in the action of the theory start to dominate. What we found is the opposite behavior --- deviations starts to dominate in low-energy dynamics (where a new de Sitter solution appears), while the high-energy regime is the same for TEGR and $f(T)$ gravity. Such a property leads to the suggestion that the model we have studied is oversimplified. Indeed, our considerations have been restricted by vacuum models, which are very different from the Universe we live in. In GR, vacuum approximation near a cosmological singularity is known to be relevant --- any matter content of the Universe (apart from a stiff fluid) has negligible influence on the cosmological dynamics, and the nature of the cosmological singularity is determined solely by the vacuum solutions. However, it is not a priory evident whether this property is still valid in modified gravity. The goal of the present paper is to study cosmological evolution in a flat anisotropic Universe in the presence of a perfect fluid and to determine the influence of the matter content of the Universe upon its dynamics.

\section{The equations of motion}
~~~~We consider cosmological models with the action of $f(T)$ theory with matter
\begin{equation}
S=\frac{1}{2K}\int e~ f(T)d^{4}x +S_m,
\label{action}
\end{equation}
where $e=\text{det}(e^A_{\mu})=\sqrt{-g}$ is the determinant consisting of the tetrad components $e^A_{\mu}$, ~~$f(T)$ is a general differentiable function of the torsion scalar $T$, ~~$S_m$ is the matter action and $K=8\pi G$. Units $\hbar=c=1$ will be used.  

    The following diagonal tetrad is chosen: 
\begin{equation}
\label{tetrad}
e^A_\mu = \mathrm{diag}(1, a(t), b(t), c(t)),
\end{equation}    
which relates to the Bianchi I metric ~~$\mathrm{d}s^2=\mathrm{d}t^2-a^2(t)\mathrm{d}x^2-b^2(t)\mathrm{d}y^2-c^2(t)\mathrm{d}z^2$,~~ where ~~$a(t)$, ~~$b(t)$,~~ $c(t)$~~ are scale factors. The torsion scalar for the chosen tetrad (\ref{tetrad}) is
\begin{equation}
\label{THabc}
T=-2(H_a H_b+H_a H_c+H_b H_c),
\end{equation}
where $H_a\equiv\frac{\dot a}{a}$, ~~$H_b\equiv\frac{\dot b}{b}$, ~~$H_c\equiv\frac{\dot c}{c}$ are anisotropic Hubble parameters, and a dot denotes the derivative with respect to time. Equation (\ref{THabc}) reduces to $T=-6H^2$ in the isotropic case $a(t)=b(t)=c(t)$, ~~$H_a=H_b=H_c=H$. Then the time derivative of torsion scalar has the
form
\begin{equation}
\label{TtHabct}
\dot T=-2(\dot H_a (H_b+H_c)+\dot H_b(H_a+H_c)+\dot H_c(H_a+H_b)),
\end{equation}

    We derive the equations of motion varying the action (\ref{action}) with respect to the chosen tetrad (\ref{tetrad}) (see, for example, \cite{anis1})
\begin{equation}
\label{constraint}
f(T)-2 T f_T=2 K\rho,
\end{equation}
\begin{equation}
\label{system1}
\dot T f_{TT}(H_b+H_c)+\frac{f}{2}+f_T\left( \dot H_b+\dot H_c +{(H_b)}^2+{(H_c)}^2+2 H_b H_c+H_a H_b+H_a H_c\right)=-Kw\rho, 
\end{equation}
\begin{equation}
\label{system2}
\dot T f_{TT}(H_a+H_c)+\frac{f}{2}+f_T\left( \dot H_a+\dot H_c +{(H_a)}^2+{(H_c)}^2+2 H_a H_c+H_a H_b+H_b H_c\right)=-Kw\rho,
\end{equation}
\begin{equation}
\label{system3}
\dot T f_{TT}(H_a+H_b)+\frac{f}{2}+f_T\left( \dot H_a+\dot H_b +{(H_a)}^2+{(H_b)}^2+2 H_a H_b+H_a H_c+H_b H_c\right)=-Kw\rho,
\end{equation}
where $\rho$ is the energy density of matter, $p$ the pressure of matter, $p=w\rho$ the matter equation of state; $w\in[-1; 1]$ is a constant. Here we denote $f_T=\frac{d f(T)}{d T}$, ~~ $f_{TT}=\frac{d^2 f(T)}{dT^2}$. 

    The continuity equation for matter is
\begin{equation}
\label{continuity}
\dot\rho+(H_a+H_b+H_c)(1+w)\rho=0.
\end{equation}

    We subtract (\ref{system1}) from the sum of (\ref{system2}) and (\ref{system3}), substitute the constraint $K\rho=\frac{f}{2}-T f_T$ and (\ref{THabc}) to the obtained equation and find
\begin{equation}
\label{system11}
H_a\dot T f_{TT}+f_T\left(\dot H_a+{H_a}^2-H_b H_c\right)=-\frac{1}{2}K(w+1)\rho.
\end{equation}  
The following two equations are obtained analogously: 
\begin{equation}
\label{system21}
H_b\dot T f_{TT}+f_T\left(\dot H_b+{H_b}^2-H_a H_c\right)=-\frac{1}{2}K(w+1)\rho, 
\end{equation}   
\begin{equation}
\label{system31}
H_c\dot T f_{TT}+f_T\left(\dot H_c+{H_c}^2-H_a H_b \right)=-\frac{1}{2}K(w+1)\rho. 
\end{equation}  
The sum of these three equations is
\begin{equation}
\label{sumequation}
(H_a+H_b+H_c)\dot T f_{TT}+f_T\left( (\dot H_a+\dot H_b +\dot H_c)+{(H_a+H_b+H_c)}^2+\frac{3}{2}T\right)=-\frac{3}{2}K(w+1)\rho.
\end{equation}

\section{The linear relation between $H_a$, $H_b$, $H_c$}
~~~~The system under investigation contains four variables --- three Hubble parameters and the matter energy density. We have also one constraint equation (\ref{constraint}), so we could expect that the number of independent variables is equal to three. However, the special very symmetric nature of equations of motion induces one more relation between the Hubble parameters. 

    To show this we subtract in pairs the equations of the system (\ref{system11})-(\ref{system31}) and find
\begin{equation}
\label{system12}
(H_b-H_a)\dot T f_{TT}+f_T\left( \dot H_b-\dot H_a+{H_b}^2-{H_a}^2+H_c(H_b-H_a)\right)=0, 
\end{equation}   
\begin{equation}
\label{system22}
(H_c-H_b)\dot T f_{TT}+f_T\left( \dot H_c-\dot H_b+{H_c}^2-{H_b}^2+H_a(H_c-H_b)\right)=0, 
\end{equation}   
\begin{equation}
\label{system32}
(H_a-H_c)\dot T f_{TT}+f_T\left( \dot H_a-\dot H_c+{H_a}^2-{H_c}^2+H_b(H_a-H_c)\right)=0. 
\end{equation}
Now we add Eq.~(\ref{system12}) multiplied by $(H_c-H_b)$ to Eq.~(\ref{system22}) multiplied by $(H_a-H_b)$
\begin{equation}
\label{system13}
2f_T\left( (\dot H_b-\dot H_a)(H_c-H_b)+(\dot H_c-\dot H_b)(H_a-H_b)\right)=0. 
\end{equation}
Analogously to (\ref{system13}) we obtain 
\begin{equation}
\label{system23}
2f_T\left( (\dot H_c-\dot H_b)(H_a-H_c)+(\dot H_a-\dot H_c)(H_b-H_c)\right)=0,
\end{equation}
\begin{equation}
\label{system33}
2f_T\left( (\dot H_a-\dot H_c)(H_b-H_a)+(\dot H_b-\dot H_a)(H_c-H_a)\right)=0.
\end{equation}
The following expression is found from these system for $f_T\neq0$:
\begin{equation}
\label{Hct}
\dot H_c(H_a-H_b)=\dot H_a(H_c-H_b)+\dot H_b(H_a-H_c).
\end{equation}
For $f_T\neq0$, $H_a\neq H_b\neq H_c$ the system (\ref{system13})-(\ref{system33}) gives us
\begin{equation}
\label{dlnHabc}
\begin{array}{l}
\frac{\dot H_b-\dot H_a}{H_b-H_a}=\frac{\dot H_c-\dot H_b}{H_c-H_b}=\frac{\dot H_a-\dot H_c}{H_a-H_c} 
~~\Rightarrow\\
\Rightarrow~~ \frac{d\Big{(}\ln(H_a-H_b)\Big{)}}{dt}=\frac{d\Big{(}\ln(H_c-H_b)\Big{)}}{dt}=\frac{d\Big{(}\ln(H_a-H_c)\Big{)}}{dt}.
\end{array}
\end{equation}
Solving these differential equations we find
\begin{equation}
\label{HH1}
H_b-H_a=C_1(H_c-H_b),
\end{equation}   
\begin{equation}
\label{HH2}
H_c-H_b=C_2(H_a-H_c),
\end{equation}   
\begin{equation}
\label{HH3}
H_b-H_a=C_3(H_a-H_c),
\end{equation}
where $C_3=C_1C_2$ is obtained after the substitution (\ref{HH2}) to (\ref{HH1}). The sum (\ref{HH2}) and (\ref{HH3}) is $H_c-H_a=(C_2+C_1C_2)(H_a-H_c)$. Then $C_2+C_1C_2=-1 ~~\Rightarrow~~ C_2=-\frac{1}{1+C_1}$ and $C_3=-\frac{C_1}{1+C_1}$. Therefore, $H_c$ is the linear combination of $H_a$ and $H_b$ for the assumptions $f_T\neq0$ and $H_a\neq H_b\neq H_c$:
\begin{equation}
\label{Hc}
H_c=-\frac{1}{C_1}H_a+\frac{1+C_1}{C_1}H_b.
\end{equation} 

\section{Dynamical system for the model $f(T)=T+f_0 T^N$}
~~~~In what follows we consider cosmological models with the Lagrangian density function $f(T)=T+f_0 T^N$, where $f_0$, ~~$N>0$ are parameters. Then the constraint (\ref{constraint}) and others field equations (\ref{system11})-(\ref{system31}) have the form
 \begin{equation} 
\label{constraint1}
\begin{array}{l}
T+f_0T^N-2T-2f_0NT^N=2K\rho ~~\Rightarrow\\
\Rightarrow~~ -T+f_0(1-2N)T^N=2K\rho,
\end{array}
\end{equation}
\begin{equation}
\label{system14}
H_a\dot T f_0 N(N-1)T^{N-2}+(1+f_0NT^{N-1})\left(\dot H_a+{H_a}^2-H_b H_c\right)=-\frac{1}{2}K(w+1)\rho, 
\end{equation}   
\begin{equation}
\label{system24}
H_b\dot T f_0 N(N-1)T^{N-2}+(1+f_0NT^{N-1})\left(\dot H_b+{H_b}^2-H_a H_c\right)=-\frac{1}{2}K(w+1)\rho, 
\end{equation}   
\begin{equation}
\label{system34}
H_c\dot T f_0 N(N-1)T^{N-2}+(1+f_0NT^{N-1})\left(\dot H_c+{H_c}^2-H_a H_b \right)=-\frac{1}{2}K(w+1)\rho. 
\end{equation}  
We find the sum of Eqs.~(\ref{system14})-(\ref{system34}) to be
\begin{equation}
\begin{array}{l}
\label{sumequation1}
(H_a+H_b+H_c)\dot T N(N-1)T^{N-2}+
\\+(1+f_0NT^{N-1})\left( (\dot H_a+\dot H_b +\dot H_c)+{(H_a+H_b+H_c)}^2+\frac{3}{2}T\right)=-\frac{3}{2}K(w+1)\rho.
\end{array}
\end{equation} 

    Due to the linear relation (\ref{Hc}) we can expect that two variables are enough for the corresponding dynamical system. For such variables we choose the torsion scalar and the sum of the three Hubble parameters; therefore, we shall use Eqs.~(\ref{constraint1}) and (\ref{sumequation1}) in the present section. Separate behavior of Hubble parameters will be considered later in the  following sections.
 
    New expansion-normalized variables are introduced as follows:
\begin{equation}
\label{xyr}
x=\frac{T}{X^2},~~~~
y=\frac{f_0(1-2N)T^{N-1}-1}{f_T},~~~~
r=\frac{K \rho}{X^2f_T},
\end{equation}
where we denote $X=H_a+H_b+H_c$. 

    The variable $y$ depends on $T^{N-1}$, therefore we can express $T^{N-1}$ and $f_T$ through $y$:
\begin{equation}
\begin{array}{l}
\label{yT}
y=\frac{f_0(1-2N)T^{N-1}-1}{1+f_0NT^{N-1}} ~~\Rightarrow~~ f_0T^{N-1}=-\frac{1+y}{2N-1+Ny},
\end{array}
\end{equation} 
then
\begin{equation}
\label{yfT}
f_T=1+f_0NT^{N-1}=1-N\frac{1+y}{2N-1+Ny}=\frac{N-1}{2N-1+Ny},
\end{equation}
\begin{equation}
\label{yfTT}
f_{TT}=f_0N(N-1)T^{N-2}=-\frac{N(N-1)(1+y)}{T(2N-1+Ny)}
\end{equation}
and
\begin{equation}
\label{yfTfTT}
\frac{f_{TT}}{f_{T}}=-\frac{N(1+y)}{T}.
\end{equation}
It follows from the definition of variable $y$ that 
\\\textbf{1).} if $y\to0$ then $T^{N-1}\to const=\frac{1}{f_0(1-2N)}$, $N\neq\frac{1}{2}$,
\\\textbf{2).} if $y\to-1$ then $T\to 0$,
\\\textbf{3).} if $y\to(1-2N)/N$ then $T\to \pm\infty$.

    Dividing the constraint equation (\ref{constraint1}) by $X^2f_T$ we find
\begin{equation} 
\label{constraintxyr}
\frac{T}{X^2}\left( \frac{-1+f_0(1-2N)T^{N-1}}{f_T}\right) =\frac{2K\rho}{X^2f_T} ~~\Rightarrow~~ xy=2r,
\end{equation} 
Therefore, the variable $r$ is not independent and can be excluded. Now we divide (\ref{sumequation}) by $X^2f_T$ and obtain   
\begin{equation}
\label{sumequationxyr}
\frac{\dot T f_{TT}}{Xf_T}+\frac{\dot X}{X^2}+1+\frac{3}{2}x=-\frac{3}{2}(w+1)r.
\end{equation}

    Taking the time derivative of (\ref{constraint1}) we have
\begin{equation} 
\begin{array}{l}
\label{dtconstraint}
-\dot T+f_0 N(1-2N)\dot T T^{N-1}=2K\dot\rho ~~\Rightarrow\\
\Rightarrow~~ -\dot T(1-f_0 N(1-2N)T^{N-1})=-2K(1+w)X\rho.
\end{array}
\end{equation}
Using (\ref{yT}), (\ref{yfT}), (\ref{constraintxyr}) and (\ref{dtconstraint}) we obtain the expression for the quantities $\frac{\dot T}{X^3}$ and $\frac{\dot T}{TX}$:
\begin{equation} 
\label{dTX3}
-\frac{\dot T}{X^3}\left( \frac{1-f_0 N(1-2N)T^{N-1}}{f_T}\right) =-2\frac{K(1+w)\rho}{X^2f_T} 
~~\Rightarrow~~ \frac{\dot T}{X^3} =\frac{xy(1+w)}{1-2N(y+1)},
\end{equation}    
and
\begin{equation} 
\label{dTXT}
\frac{\dot T}{X^3}=\frac{\dot T}{TX}\frac{T}{X^2} =\frac{xy(1+w)}{1-2N(y+1)} 
~~\Rightarrow~~ \frac{\dot T}{TX}=\frac{y(1+w)}{1-2N(y+1)}.
\end{equation}
Taking into account (\ref{yfTfTT}) we have
\begin{equation}
\begin{array}{l}
\label{dXX2}
-\frac{\dot T }{TX}N(1+y)+\frac{\dot X}{X^2}+1+\frac{3}{2}x=-\frac{3}{4}(w+1)xy ~~\Rightarrow\\
\\\Rightarrow~~ \frac{\dot X}{X^2}=-1-\frac{3}{2}x+(w+1)y\left( \frac{3x}{4}+\frac{N(1+y)}{1-2N(y+1)}\right).
\end{array}
\end{equation}
   We differentiate the new variables $x$, $y$ with respect to ~~$\ln(abc)$,~~ where $d \Big{(}\ln (abc)\Big{)}=dt\left( \frac{\dot a}{a}+\frac{\dot b}{b}+\frac{\dot c}{c}\right) =Xdt$, and we find the system of first-order differential equations
\begin{equation}
\label{dx0}
\frac{dx}{d(\ln (abc))}=\frac{dx}{Xdt}=\frac{\dot T}{X^3}-2\frac{\dot X}{X^2}x,
\end{equation}
\begin{equation}
\label{dy0}
\frac{dy}{d (\ln (abc))}=\frac{dy}{Xdt}=-\frac{\dot T}{TX}(1-2N)(1+y)+\frac{\dot T}{TX}Ny(1+y).
\end{equation}
Using Eqs.~(\ref{dTX3}), (\ref{dTXT}), (\ref{dXX2}) we finally obtain
\begin{equation}
\label{dx}
\frac{dx}{d(\ln (abc))}=x(2+3x)\left( 1+\frac{1}{2}y(1+w)\right) ,
\end{equation}
\begin{equation}
\label{dy}
\frac{dy}{d (\ln (abc))}=\frac{y(1+y)(2N-1+Ny)(1+w)}{1-2N(y+1)}.
\end{equation}

\section{Stationary points}
~~~~We solve the system (\ref{dx}), (\ref{dy}), find stationary points and calculate the corresponding eigenvalues for each point in order to obtain their type of stability in the linear approach.
\\\textbf{1.} $x=0$, $y=(1-2N)/N$.
\\Eigenvalues are
\begin{equation}
\begin{array}{l}
\lambda_1=(1+w(1-2N))/N>0~~ \text{for $N>1, w\in\left[ -1; \frac{1}{2N-1}\right)$}\\
~~~~~~~~~~~~~~~~~~~~~~~~~~~~~~~~~<0~~ \text{for $N>1, w\in\left(\frac{1}{2N-1}; 1\right]$},\\
\\\lambda_2=(1+w)(N-1)/N\geqslant0~~ \text{for $N>1$, $w\in[-1; 1]$}.
\end{array}
\end{equation}
This stationary point is either an unstable node for $N>1$, $w\in\left( -1; \frac{1}{2N-1}\right)$ or a saddle point for $N>1$, $w\in\left(\frac{1}{2N-1}; 1\right]$. For $w=\frac{1}{2N-1}$, $N>1$, we find $\lambda_1=0$, $\lambda_2>0$ and this fixed point is unstable.
\\
\\\textbf{2.} $x=-2/3$, $y=(1-2N)/N$.
\\The eigenvalues are calculated to be
\begin{equation}
\begin{array}{l}
\lambda_1=(w(2N-1)-1)/N>0~~ \text{for $N>1, w\in\left(\frac{1}{2N-1}; 1\right]$}\\
~~~~~~~~~~~~~~~~~~~~~~~~~~~~~~~~~<0~~ \text{for $N>1, w\in\left[ -1; \frac{1}{2N-1}\right)$},\\
\\\lambda_2=(1+w)(N-1)/N\geqslant0~~ \text{for $N>1$, $w\in[-1; 1]$}.
\end{array}
\end{equation}
We see from the sings of the eigenvalues that this stationary point is either an unstable node for $N>1, w\in\left(\frac{1}{2N-1}; 1\right]$ or a saddle for $N>1, w\in\left(-1; \frac{1}{2N-1}\right)$. For $w=\frac{1}{2N-1}$, $N>1$, this point is unstable as $\lambda_2>0$.  
\\
\\\textbf{3.} $x=-2/3$, $y=-1$.
\\We find the eigenvalues
\begin{equation}
\begin{array}{l}
\lambda_1=w-1\leqslant0~~ \text{for $w\in[-1; 1]$},\\
\lambda_2=(1+w)(1-N)\leqslant0~~ \text{for $N>1$, $w\in[-1; 1]$}.
\end{array}
\end{equation}
It is a stable node for $N>1$, $w\in(-1; 1)$.    
\\
\\\textbf{4.} $x=-2/3$, $y=0$.
\\The eigenvalues are obtained:
\begin{equation}
\begin{array}{l}
\lambda_1=-2<0,\\
\lambda_2=-1-w\leqslant0~~ \text{for $w\in[-1; 1]$},
\end{array}
\end{equation}
which indicate that it is a stable node for $w\in(-1; 1]$.

     Using stationary point coordinates we calculate $\frac{\dot X}{X^2}=0$,  $\frac{\dot T}{TX}=0$ and find $X(t)$, $T(t)$, $\rho(t)$,
\begin{equation}
\label{X4}
X(t)=X_0,
\end{equation}
\begin{equation}
\label{T4}
T(t)=T_0,
\end{equation}
\begin{equation}
\label{rho4}
\rho(t)=\rho_0{e}^{-X_0(1+w)(t-t_0)}.
\end{equation}
For $y=0$ we find ${T_0}^{N-1}=\frac{1}{f_0(1-2N)}$. As in this solution the sum of Hubble parameters $X$ and the torsion scalar $T$ are constants, then all Hubble parameters are equal, $H_a=H_b=H_c=H_0=\pm\sqrt{-\frac{T_0}{6}}$. This is the de Sitter solution. 
\\
\\\textbf{5.} In partial case of $w=w_{cr}=1/(2N-1)$ the stationary line exists: 
\\$x\in(-\infty; +\infty)$, $y=(1-2N)/N$.
\\The eigenvalues for this stationary line are 
\begin{equation}
\begin{array}{l}
\lambda_1=0,\\
\lambda_2=2(N-1)/(2N-1)>0~~ \text{for $N>1$}.
\end{array}
\end{equation}
As $\lambda_2>0$ for $N>1$, this stationary line is unstable.

\section{The asymptotic power-law solutions}
~~~~Since most of known exact or asymptotic solutions in homogeneous cosmology have a power-law form, we check the existence of the asymptotic power-law solution of the form 
\begin{equation}
\label{HaHbHc}
H_a=\frac{p_1}{t-t_0},~~~~ H_b=\frac{p_2}{t-t_0},~~~~ H_c=\frac{p_3}{t-t_0}.
\end{equation}
Then the torsion scalar $T$, its time derivative $\dot T$, the energy density of matter $\rho$ are
\begin{equation}
\label{THabct}
T=-\frac{2}{{(t-t_0)}^2}(p_1p_2+p_1p_3+p_2p_3),
\end{equation}    
\begin{equation}
\label{dotTHabct}
\dot T=\frac{4}{{(t-t_0)}^3}(p_1p_2+p_1p_3+p_2p_3),
\end{equation}   
\begin{equation}
\label{rhot}
\rho=\rho_0 {(t-t_0)}^{-(w+1)(p_1+p_2+p_3)}.  
\end{equation}
We substitute the solution (\ref{HaHbHc}) into the initial system (\ref{constraint1})-(\ref{system34}) for two asymptotic limits: \textbf{1).}~$t\to t_0$,~~\textbf{2).}~$t\to+\infty$. We shall consider only the case of $N>1$ in all calculations below.
\\
\\\textbf{1).} For $t\to t_0$ the torsion scalar $|T|\to\infty$ and $|T^N|\gg |T|$.
Then the term $-T$ can be neglected in comparison with $f_0(1-2N)T^N$ in the constraint (\ref{constraint1}) 
\begin{equation}
\label{constraint2}
2K\rho= f_0(1-2N)T^N,
\end{equation}  
and using Eq.~(\ref{rhot}) we find
\begin{equation}
\label{constraint3}
2K\rho_0 {(t-t_0)}^{-(w+1)(p_1+p_2+p_3)}= \frac{f_0}{{(t-t_0)}^{2N}}(1-2N){(-2)}^N{\left( p_1p_2+p_1p_3+p_2p_3\right) }^N.
\end{equation}  
We equate power indices and multipliers in Eq.~(\ref{constraint3}) and obtain
\begin{equation}
\label{1p1p2p3}
p_1+p_2+p_3=\frac{2N}{w+1},~~ w\neq-1,
\end{equation}
\begin{equation}
\label{1rho0}
\rho_0 = \frac{f_0}{2K}(1-2N){(-2)}^N{\left( p_1p_2+p_1p_3+p_2p_3\right) }^N.
\end{equation}
    Substituting the power-law solution (\ref{HaHbHc}) into Eq.~(\ref{sumequation1}) we obtain
\begin{equation}
\label{1sump}
{p_1}^2+{p_2}^2+{p_3}^2=(2N-1)(p_1+p_2+p_3)+\left( 1+\frac{3(w+1)(1-2N)}{2N}\right) (p_1p_2+p_1p_3+p_2p_3).
\end{equation} 
Then we rewrite the substitution of the solution (\ref{HaHbHc}) into the system (\ref{system14})-(\ref{system34}) and subtract in pairs the obtained equations,
\begin{equation}
\label{system15}
(p_1-p_2)\Big{(} 2(N-1)+1-(p_1+p_2+p_3)\Big{)} =0, 
\end{equation}   
\begin{equation}
\label{system25}
(p_1-p_3)\Big{(} 2(N-1)+1-(p_1+p_3+p_2)\Big{)} =0, 
\end{equation}   
\begin{equation}
\label{system35}
(p_2-p_3)\Big{(} 2(N-1)+1-(p_2+p_3+p_1)\Big{)} =0.
\end{equation}
   Therefore, we have two asymptotic power-law solutions for $t\to t_0$:
\\\textbf{a).} Isotropic asymptotic regime with $p_1=p_2=p_3=p$,
\begin{equation}
\label{p1}
p=\frac{2N}{3(w+1)},~~ w\in(-1; 1],
\end{equation}
\begin{equation}
\label{rho0p1}
\rho_0 = \frac{f_0}{2K}(1-2N){(-2)}^N{\left( 3p^2\right) }^N,
\end{equation}
where we have used (\ref{1p1p2p3}), (\ref{1rho0}).
\\
\\\textbf{b).} Anisotropic asymptotic regime with $p_1+p_2+p_3=2N-1$,
\begin{equation}
\label{w}
w=w_{cr}=\frac{1}{2N-1}, 
\end{equation}
\begin{equation}
\label{rho01}
\rho_0 = \frac{f_0}{2K}(1-2N){(-2)}^N{\left( p_1p_2+p_1p_3+p_2p_3\right) }^N.
\end{equation}
We find from (\ref{1sump})
\begin{equation}
\label{sump1}
{p_1}^2+{p_2}^2+{p_3}^2={(2N-1)}^2-2(p_1p_2+p_1p_3+p_2p_3).
\end{equation}
Here $p_1p_2+p_1p_3+p_2p_3\neq0$.

    The found isotropic solution \textbf{a).} corresponds to the fixed point \textbf{2} and the anisotropic one \textbf{b).} corresponds to the stationary line \textbf{5}.   
\\
\\\textbf{2).} For $t\to+\infty$ we find that $T\to 0$, $|T^N|\ll |T|$. Neglecting $f_0(1-2N)T^N$ in (\ref{constraint1}) we obtain
\begin{equation}
\label{constraint4}
2K\rho= -T.
\end{equation} 
Taking into account (\ref{rhot}) we have 
\begin{equation}
\label{constraint5}
2K\rho_0 {(t-t_0)}^{-(w+1)(p_1+p_2+p_3)}= \frac{2}{{(t-t_0)}^2}(p_1p_2+p_1p_3+p_2p_3).
\end{equation}  
Equating power indices and multipliers in (\ref{constraint5}) we obtain
\begin{equation}
\label{2p1p2p3}
p_1+p_2+p_3=\frac{2}{w+1},~~ w\neq-1,
\end{equation}
\begin{equation}
\label{2rho0}
\rho_0 = \frac{1}{K}(p_1p_2+p_1p_3+p_2p_3).
\end{equation}
     The substitution (\ref{HaHbHc}) into the system (\ref{system14})-(\ref{system34}) and the subtraction in pairs of the resulting equations lead to
\begin{equation}
\label{system16}
(p_1-p_2)(-1+p_1+p_2+p_3)=0,
\end{equation}   
\begin{equation}
\label{system26}
(p_1-p_3)(-1+p_2+p_3+p_1)=0, 
\end{equation}   
\begin{equation}
\label{system36}
(p_2-p_3)(-1+p_1+p_3+p_2)=0.
\end{equation}
Thus, we have the 3-d power-law solution in a form which exists for $t\to+\infty$
\\\textbf{c).} Isotropic asymptotic regime with $p_1=p_2=p_3=p$,
\begin{equation}
\label{p2}
p=\frac{2}{3(w+1)},~~ w\in(-1; 1],
\end{equation}
where we have used (\ref{2p1p2p3}),
\begin{equation}
\label{rho0p2}
\rho_0 =\frac{3 p^2}{K}.
\end{equation}
This isotropic solution \textbf{c).} corresponds to the fixed point \textbf{3}. 

\section{More complicated asymptotic power-law solution}
~~~~In the previous section we have identified the solutions in three fixed points from our list, one more (the de Sitter solution) has been identified in the Sect.~3. However, the nature of a solution in the point \textbf{1} is still to be determined. As $X=0$ at this point, it should correspond to some high-energy regime.

    To reveal the meaning of the point \textbf{1} we try to check a more complicated form for the Hubble parameters:
\begin{equation}
\label{Ha1}     
H_a=\frac{p_{a1}}{t-t_0}+\frac{p_{a2}}{{(t-t_0)}^{\alpha-1}},
\end{equation} 
\begin{equation} 
\label{Hb1}    
H_b=\frac{p_{b1}}{t-t_0}+\frac{p_{b2}}{{(t-t_0)}^{\alpha-1}},
\end{equation}
\begin{equation}
\label{Hc1}     
H_c=\frac{p_{c1}}{t-t_0}+\frac{p_{c2}}{{(t-t_0)}^{\alpha-1}}.
\end{equation}   
Here $p_{a1}$, $p_{a2}$, $p_{b1}$, $p_{b2}$, $p_{c1}$, $p_{c2}$, $\alpha$ are constants, which satisfy the following relations:
\begin{equation}
\label{sumpa1pb1pc1}
p_{a1}+p_{b1}+p_{c1}\neq 0,
\end{equation}
\begin{equation}
\label{pa1pb1pc1}
p_{a1}p_{b1}+p_{a1}p_{c1}+p_{b1}p_{c1}=0~~\Rightarrow~~ {(p_{a1}+p_{b1}+p_{c1})}^2={p_{a1}}^2+{p_{b1}}^2+{p_{c1}}^2,
\end{equation}
\begin{equation}
\label{alpha}
-1<\alpha-1<1. 
\end{equation}
The assumptions (\ref{sumpa1pb1pc1}) and (\ref{pa1pb1pc1}) lead to $p_{a1}\neq p_{b1}\neq p_{c1}$. Equation (\ref{alpha}) shows that the first terms in (\ref{Ha1})-(\ref{Hc1}) dominate for $t\to t_0$ and the Hubble parameters increase as ${(t-t_0)}^{-1}$:
\begin{equation}
\label{Ha2}     
H_a=\frac{p_{a1}}{t-t_0}+\frac{p_{a2}}{{(t-t_0)}^{\alpha-1}}\to\frac{p_{a1}}{t-t_0}\to\pm\infty,
\end{equation} 
\begin{equation} 
\label{Hb2}    
H_b=\frac{p_{b1}}{t-t_0}+\frac{p_{b2}}{{(t-t_0)}^{\alpha-1}}\to\frac{p_{b1}}{t-t_0}\to\pm\infty,
\end{equation}
\begin{equation}
\label{Hc2}     
H_c=\frac{p_{c1}}{t-t_0}+\frac{p_{c2}}{{(t-t_0)}^{\alpha-1}}\to\frac{p_{b1}}{t-t_0}\to\pm\infty,
\end{equation}    
  
    The torsion scalar $T$ in this solution is 
\begin{equation}
\begin{array}{l}
\label{THabct1}
T=-2\left\lbrace  {(t-t_0)}^{-2}\left[ p_{a1}p_{b1}+p_{a1}p_{c1}+p_{b1}p_{c1}\right] +\right. 
\\\left. + {(t-t_0)}^{-\alpha}\left[ p_{a1}(p_{b2}+p_{c2})+p_{b1}(p_{a2}+p_{c2})+p_{c1}(p_{a2}+p_{b2})\right] +\right. 
\\\left. + {(t-t_0)}^{2-2\alpha}\left[ p_{a2}p_{b2}+p_{a2}p_{c2}+p_{b2}p_{c2}\right]\right\rbrace.
\end{array}
\end{equation} 
Taking into account (\ref{pa1pb1pc1}), (\ref{alpha}) we find from (\ref{THabct1}) for $t\to t_0$ 
\begin{equation}
\label{THabct2}
T=-2\left\lbrace {(t-t_0)}^{-\alpha}\left[ p_{a1}(p_{b2}+p_{c2})+p_{b1}(p_{a2}+p_{c2})+p_{c1}(p_{a2}+p_{b2})\right] \right\rbrace\rightarrow\pm\infty.
\end{equation} 

    Therefore, despite neglecting the second terms in the Hubble parameters $H_a$, $H_b$, $H_c$ for $t\to t_0$, these terms are important for the torsion scalar $T$.
    
    The time derivative of torsion scalar is    
\begin{equation}
\label{dotTHabct1}
\dot T=2\alpha\left\lbrace {(t-t_0)}^{-\alpha-1}\left[ p_{a1}(p_{b2}+p_{c2})+p_{b1}(p_{a2}+p_{c2})+p_{c1}(p_{a2}+p_{b2})\right] \right\rbrace \rightarrow\pm\infty.
\end{equation}   
As $|T^N|\gg |T|$ for $T\to\pm\infty$, we find from the constraint equation (\ref{constraint1})
\begin{equation}
\label{constraint6}
2K\rho= f_0(1-2N)T^N.
\end{equation}  
Substituting (\ref{Ha1})-(\ref{Hc1}) to the continuity equation we have
\begin{equation}
\label{rhot1}
\rho=\rho_0 {(t-t_0)}^{-(w+1)(p_{a1}+p_{b1}+p_{c1})}.  
\end{equation}
Then we substitute the expressions for $\rho$ and $T$ (\ref{rhot1}), (\ref{THabct2}) to (\ref{constraint6}) and find
\begin{equation}
\begin{array}{l}
\label{constraint7}
2K\rho_0 {(t-t_0)}^{-(w+1)(p_{a1}+p_{b1}+p_{c1})}=
\\= {(t-t_0)}^{-N\alpha}f_0(1-2N){(-2)}^N{\left[ p_{a1}(p_{b2}+p_{c2})+p_{b1}(p_{a2}+p_{c2})+p_{c1}(p_{a2}+p_{b2})\right]}^N.
\end{array}
\end{equation}  
Equating the power-law indices and coefficients in (\ref{constraint7}) we obtain
\begin{equation}
\label{pa1pb1pc11}
p_{a1}+p_{b1}+p_{c1}=\frac{N\alpha}{w+1},~~ w\neq-1,
\end{equation}
\begin{equation}
\label{rho02}
\rho_0 = \frac{f_0}{2K}(1-2N){(-2)}^N{\left[ p_{a1}(p_{b2}+p_{c2})+p_{b1}(p_{a2}+p_{c2})+p_{c1}(p_{a2}+p_{b2})\right]}^N.
\end{equation}
  
    We substitute complex power-law solution to the system (\ref{system14})-(\ref{system34}), then subtract in pairs the obtained equations and find
\begin{equation}
\label{system17}
N(p_{a1}-p_{b1})\Big{(} -1+p_{a1}+p_{b1}+p_{c1}-\alpha(N-1)\Big{)} =0,
\end{equation}   
\begin{equation}
\label{system27}
N(p_{a1}-p_{c1})\Big{(} -1+p_{a1}+p_{b1}+p_{c1}-\alpha(N-1)\Big{)} =0,
\end{equation}   
\begin{equation}
\label{system37}
N(p_{b1}-p_{c1})\Big{(} -1+p_{a1}+p_{b1}+p_{c1}-\alpha(N-1)\Big{)} =0.
\end{equation} 
Here we keep only dominating terms with ${(t-t_0)}^{-\alpha(N-1)+2}$. As our initial assumptions lead to $p_{a1}\neq p_{b1}\neq p_{c1}$, the factor $-1+p_{a1}+p_{b1}+p_{c1}-\alpha(N-1)$ in the system (\ref{system17})-(\ref{system37})
should be equal to zero, which allows us to find $\alpha$:
\begin{equation}
\label{alpha1}
\alpha=\frac{p_{a1}+p_{b1}+p_{c1}-1}{N-1}, ~~N\neq 1.
\end{equation} 
Taking into account (\ref{pa1pb1pc11}) we have
\begin{equation}
\label{alpha2}
\alpha=\frac{1}{N-1}\left( \frac{N\alpha}{w+1}-1\right)~~\Rightarrow~~\alpha=\frac{1+w}{1-w(N-1)},
\end{equation}
where $\alpha$ satisfies the initial assumption (\ref{alpha}) $-1<\alpha-1<1$ ~~for~~ $-1<w<\frac{1}{2N-1}$.

    Therefore, the complex power-law solution exists with the following coefficients:
\begin{equation}
\label{pa1pb1pc12}
p_{a1}+p_{b1}+p_{c1}=\frac{N}{1-w(N-1)},~~ w\in\Big{(}-1; \frac{1}{2N-1}\Big{)},
\end{equation}
\begin{equation}
\label{alpha3}
\alpha=\frac{1+w}{1-w(N-1)},
\end{equation}
\begin{equation}
\label{rho03}
\rho_0 = \frac{f_0}{2K}(1-2N){(-2)}^N{\left[ p_{a1}(p_{b2}+p_{c2})+p_{b1}(p_{a2}+p_{c2})+p_{c1}(p_{a2}+p_{b2})\right]}^N,
\end{equation}
\begin{equation}
\label{sump5}
{p_{a1}}^2+{p_{b1}}^2+{p_{c1}}^2={(p_{a1}+p_{b1}+p_{c1})}^2.
\end{equation}
Here $p_{a1}p_{b1}+p_{a1}p_{c1}+p_{b1}p_{c1}=0$. The found solution corresponds to the stationary point \textbf{1}.

\section{Numerical examples of cosmological scenarios}
\begin{figure}[hbtp]
\includegraphics[scale=0.44]{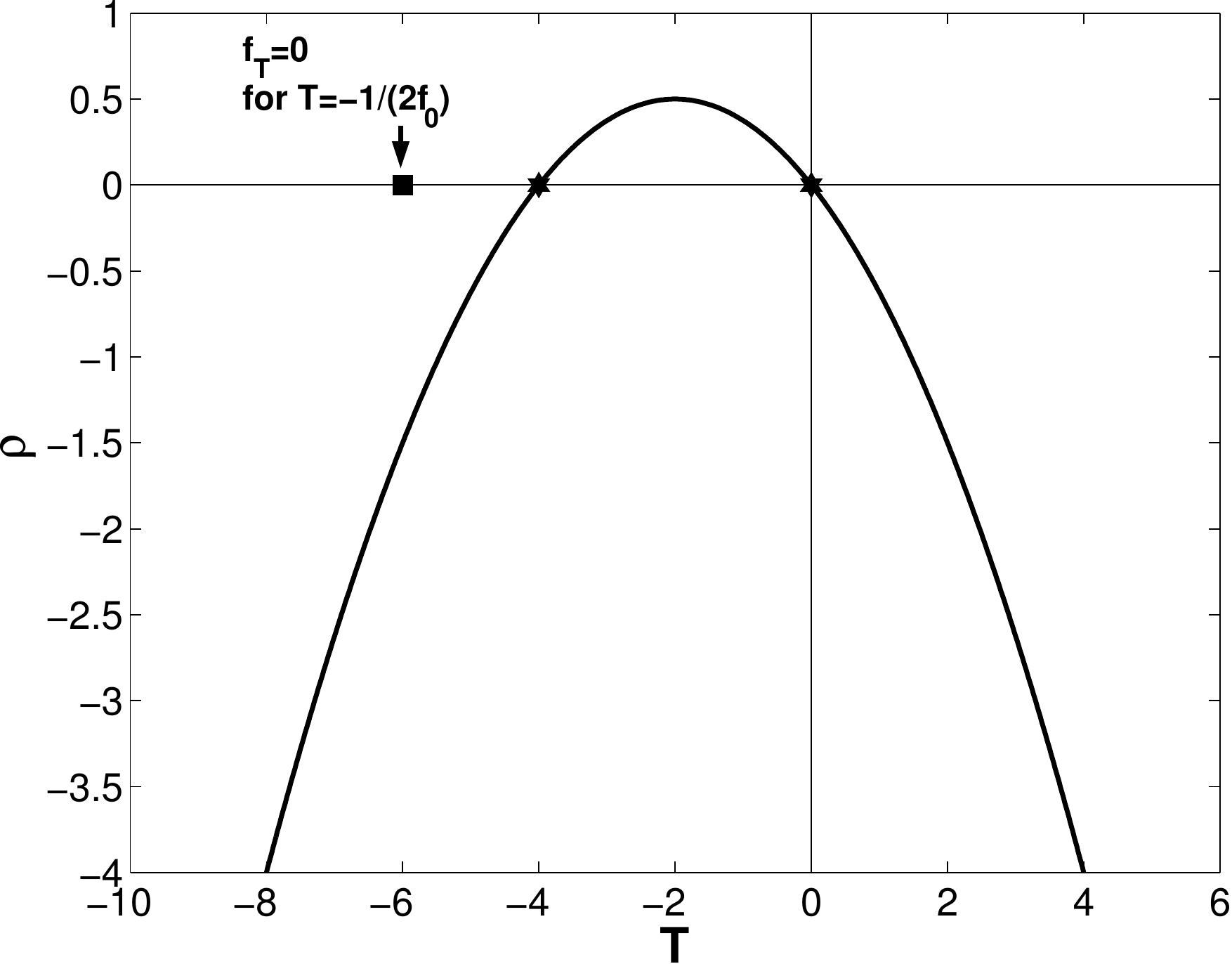}\qquad
\includegraphics[scale=0.44]{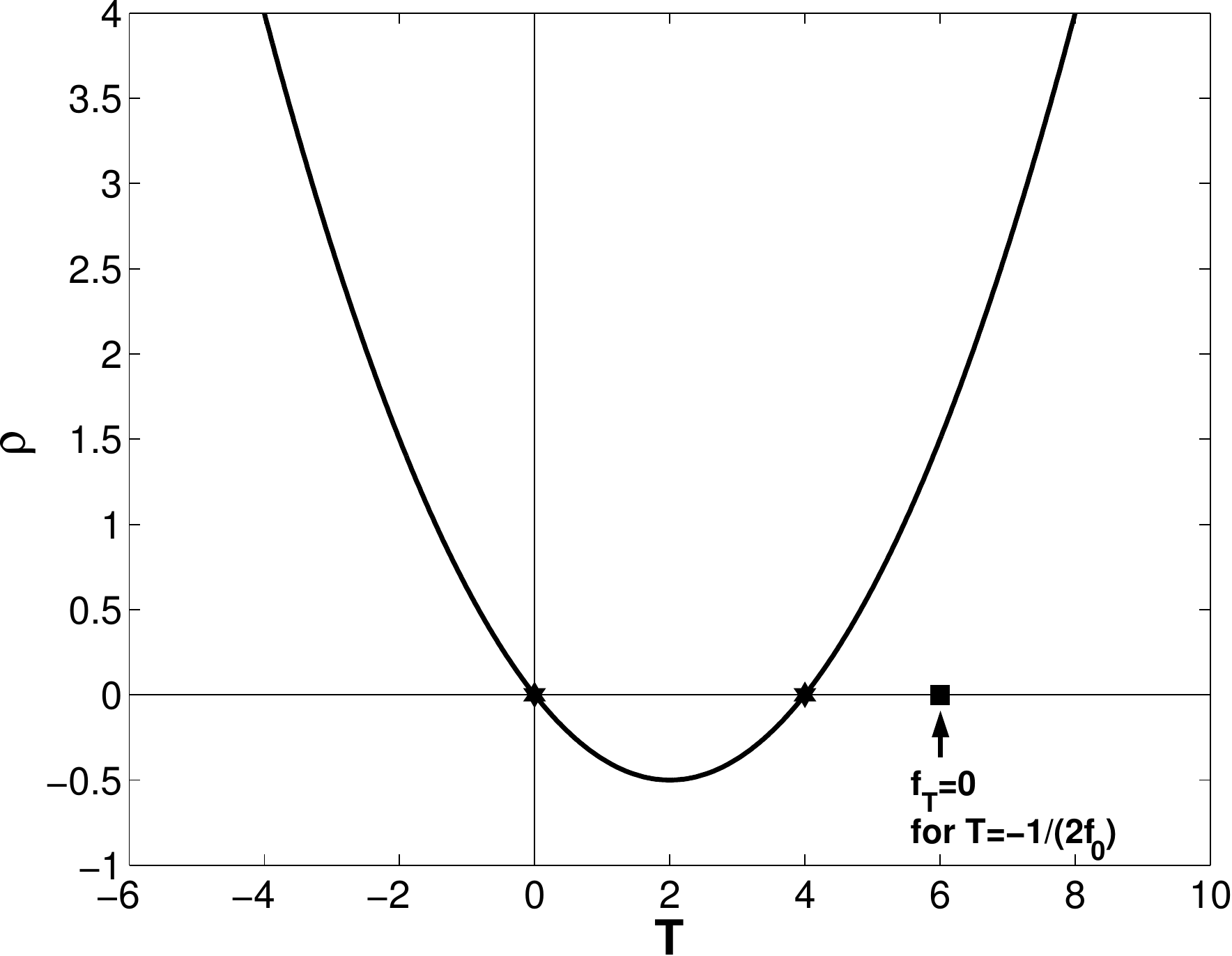} 
\caption{The dependence $\rho(T)=\frac{-T-3f_0T^2}{2K}$, which follows from the constraint equation (\ref{constraint1}) for $N=2$. The parameter $f_0=\frac{1}{12}$ for the left picture, $f_0=-\frac{1}{12}$ for the right plot and for both graphs $K=1$.}
\label{Fig1}
\end{figure}

~~~~Fixed points analysis can get information only about local properties of the dynamics. In the present section we consider global properties for models $f(T)=T+f_0 T^N$ with integer~$N$. Since they are different for even and odd $N$, we consider these cases separately, choosing $N=2$ and $N=3$ for numerical studies. First of all, the constraint equation imposes limitations on the possible values of $T$. In the case of $f(T)=T+f_0T^2$ the constraint equation~(\ref{constraint1}) is a quadratic one for the torsion scalar $T$. It has two branches of solutions: $T_1$ and $T_2$. We find the intervals of allowed values of $T$ from the condition $\rho\geqslant 0$ (see Fig.~\ref{Fig1}, where the function $\rho(T)=\frac{-T+(1-2N)f_0T^N}{2K}$ is plotted for $N=2$):
\\
\\1). for $f_0>0$ the torsion scalar $T\in\left[ \frac{1}{-3f_0}; 0\right]$, where $\frac{1}{-3f_0}\leqslant T_2<\frac{1}{-6f_0}$ and $ \frac{1}{-6f_0}\leqslant T_1\leqslant 0$,
\\2). for $f_0<0$ the torsion scalar $T\in(-\infty; 0]\cup\Big{[}\frac{1}{-3f_0}; +\infty\Big{)}$, where $T_1\leqslant 0$ and $T_2\geqslant\frac{1}{-3f_0}> 0$.
\\
\\In particular, we can see that the absolute value of the torsion scalar is restricted from above for $f_0>0$. This means that the points \textbf{1}, \textbf{2} and the line~\textbf{5} cannot be realized. Numerical studies (see below) indicate that instead of an analog of the Big Bang singularity the Universe starts its evolution from a non-standard singularity where $T$ is finite (and $H_a\neq H_b\neq H_c$ are finite) while $\dot T$ diverges. Note that for the isotropic case the same conclusion (large enough $H$ being unreachable if $f_0>0$) has been already drawn in \cite{Laur}. 
    
    We investigate numerically the model $f(T)=T+f_0T^2$ using the system (\ref{Hat4})-(\ref{Hct1}) (see Appendix), which is obtained from the initial equations of motion. The following types of scenarios are realized.  

\begin{figure}[hbtp]
\includegraphics[scale=0.44]{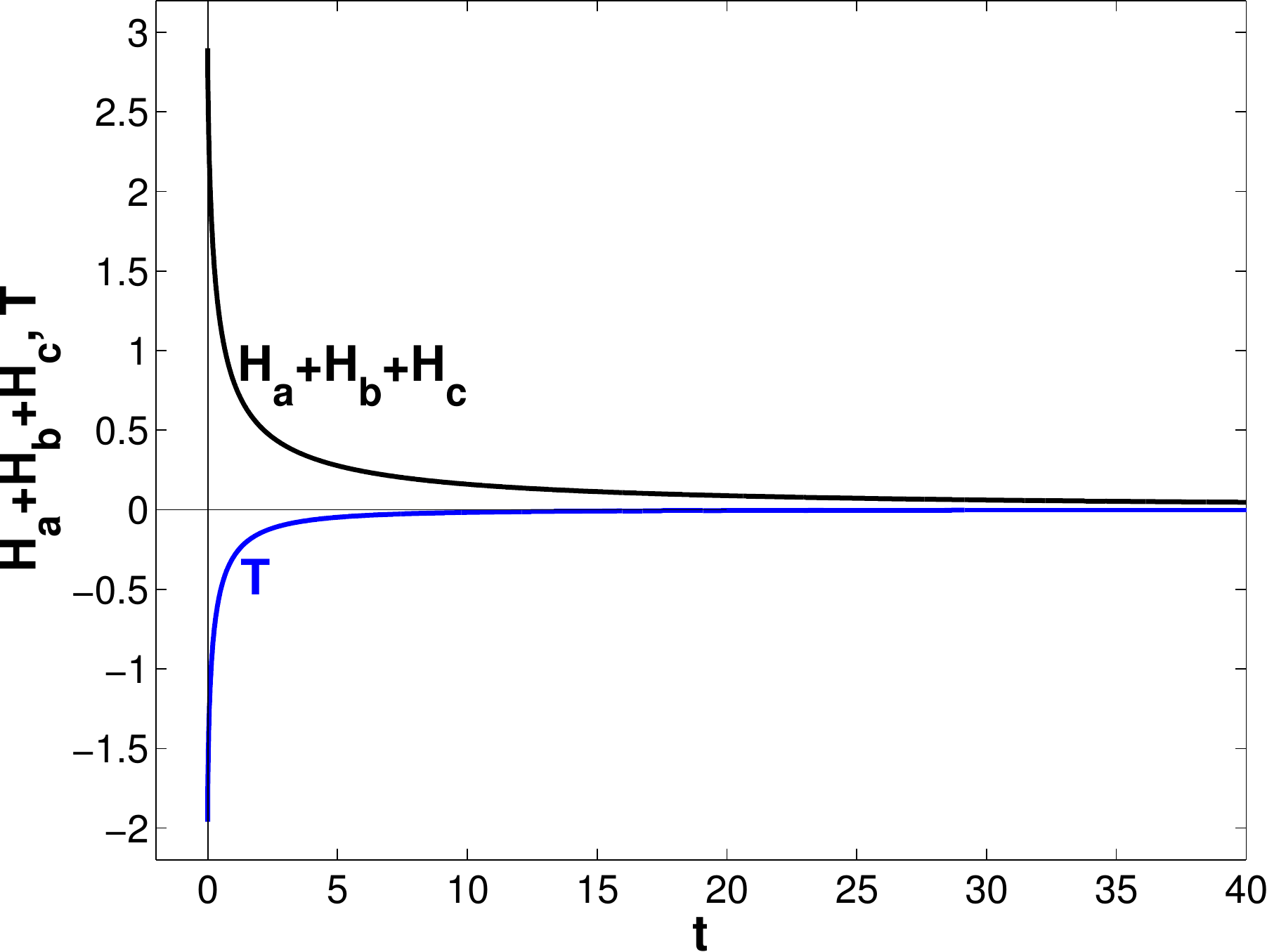}\qquad
\includegraphics[scale=0.44]{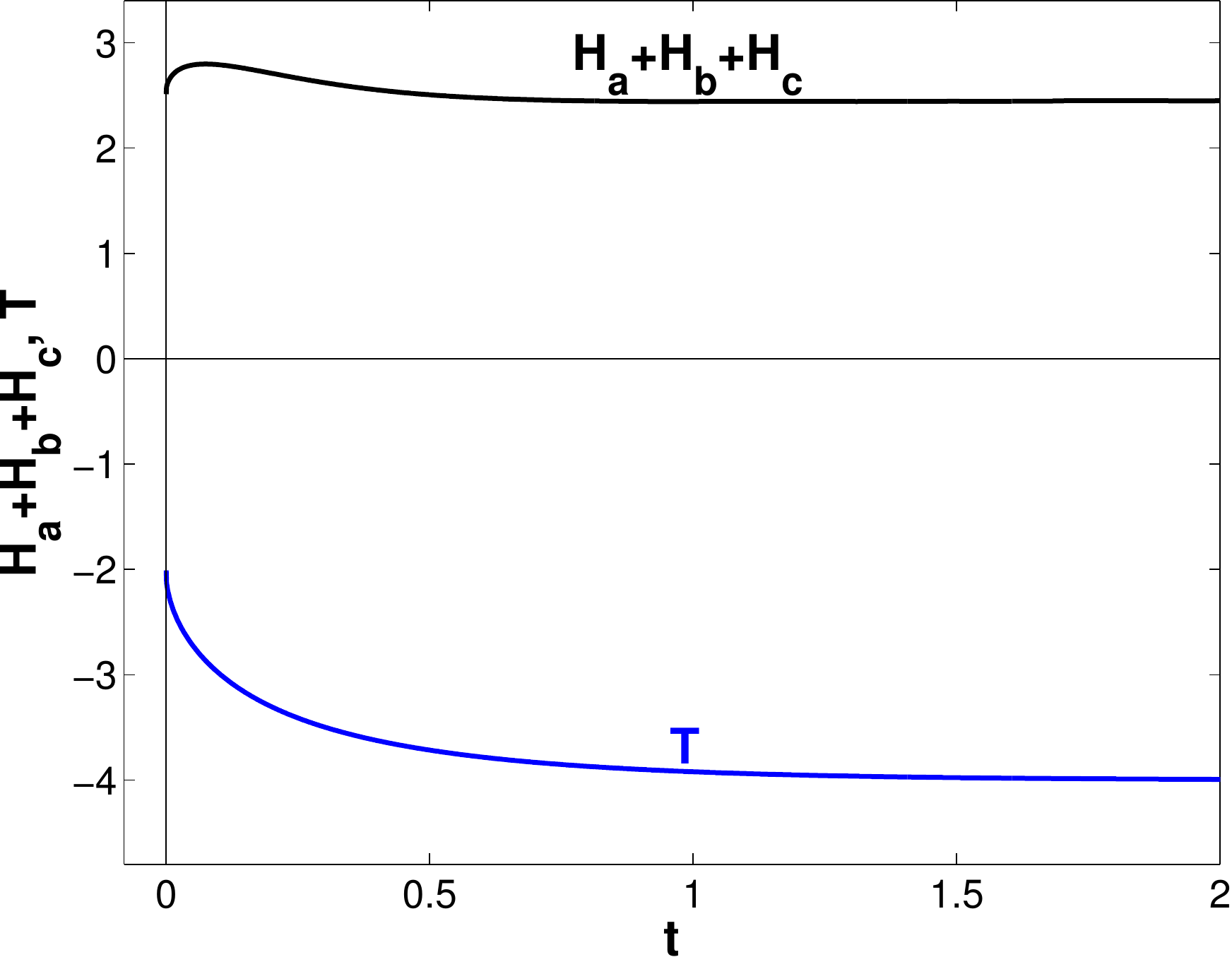} 
\caption{The evolution of the torsion scalar $T$ and sum of Hubble parameters $X=H_a+H_b+H_c$ for $N=2$, $f_0=\frac{1}{12}$, $w=0$, $K=1$. The initial values are $H_a(0)=-0.4$, $H_b(0)=1$, $H_c(0)=2.3$ for the left plot and $H_a(0)=-0.24$, $H_b(0)=0.9$, $H_c(0)=1.85$ for the right one. The left picture demonstrates the cosmological scenario \textbf{Ia} and the right graph shows the scenario \textbf{Ib}.}
\label{Fig2}
\end{figure}

    For $f_0>0$ two variants of evolution are possible:
\begin{itemize}
\item[\textbf{Ia.}] along the branch $T_1$ from the non-standard singularity with $T=\frac{1}{-6f_0}$ to the isotropic solution of the fixed point \textbf{3} (see the left plot in Fig.~\ref{Fig2});
\item[\textbf{Ib.}] along the branch $T_2$ from the non-standard singularity with $T=\frac{1}{-6f_0}$ to de Sitter solution of the fixed point \textbf{4} (see the right graph in Fig.~\ref{Fig2}).
\end{itemize}

    Note that the torsion scalar $T$ changes in a very restricted zone during a cosmological evolution along the \textbf{Ib} case --- the initial absolute value of the torsion scalar $|T_{in}|=|T_0|=\big{|}\frac{1}{-6f_0}\big{|}$ at the singularity ($T_{in}$ corresponds to the maximum of the function $\rho(T)$) is only two times less than its final absolute value $|T_{fin}|=|T_0|=\big{|}\frac{1}{-3f_0}\big{|}$ at the de Sitter point. It is easy to see that for a general power-law function $f(T)=T+f_0T^N$ the ratio of final $|T_{fin}|=|T_0|={\big{|}\frac{1}{f_0(1-2N)}\big{|}}^{\frac{1}{N-1}}$ and initial $|T_{in}|={\big{|}\frac{1}{f_0N(1-2N)}\big{|}}^{\frac{1}{N-1}}$ values of the torsion scalar is equal to $N^{\frac{1}{N-1}}$. This means that this de Sitter point cannot be related to our Universe.

\begin{figure}[hbtp]
~~~~~~~~~~~~\includegraphics[scale=0.68]{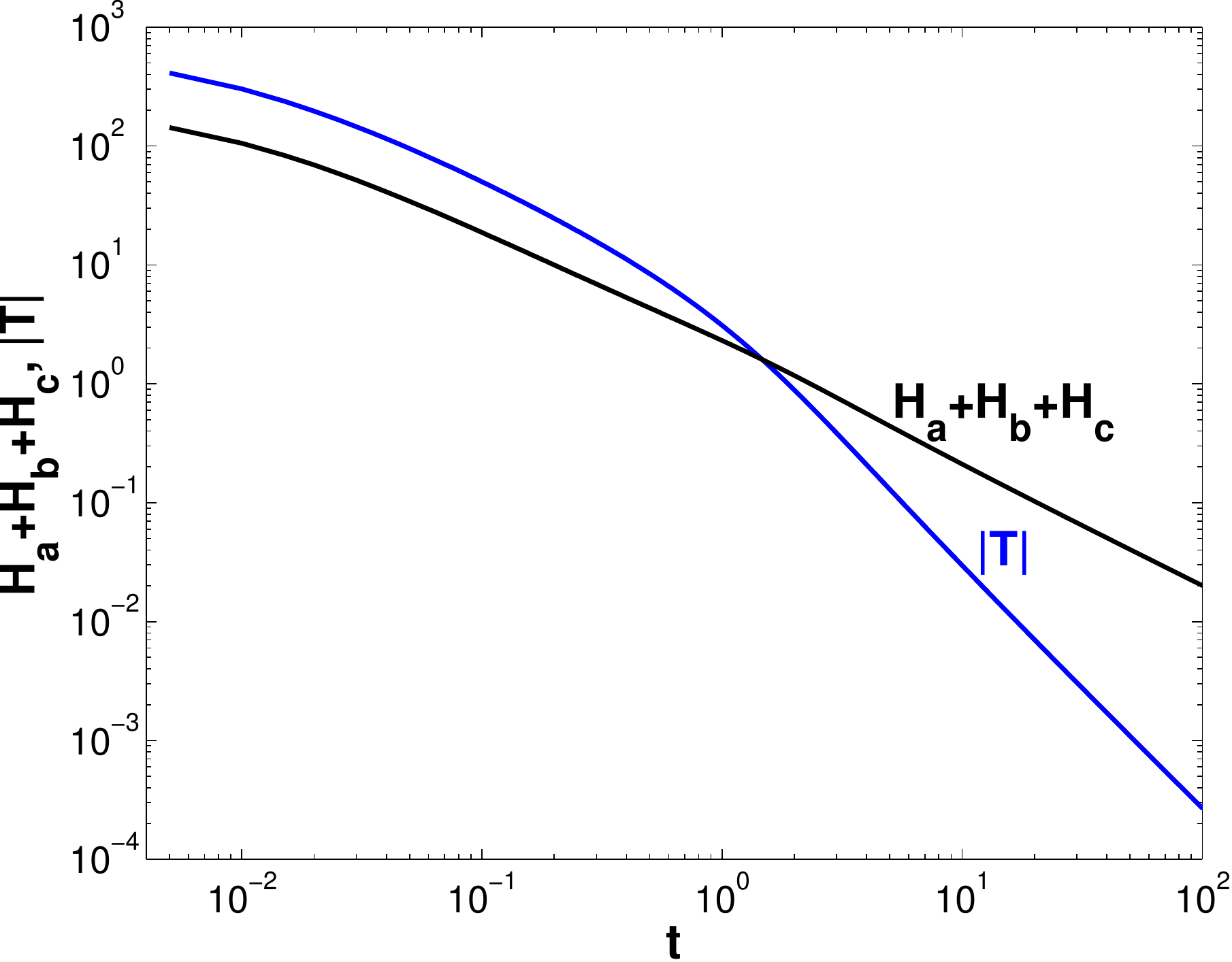}
\caption{The evolution of the torsion scalar $|T|$ and the sum of the Hubble parameters $X=H_a+H_b+H_c$ for $N=2$, $f_0=-\frac{1}{12}$, $w=0$, $K=1$. The initial values are $H_a(0)=220$, $H_b(0)=-10$, $H_c(0)=12$. This figure demonstrates the cosmological scenario \textbf{IIa}, where the Universe evolves from the anisotropic solution of point \textbf{1} to the isotropic regime of the point \textbf{3}.}
\label{Fig3}
\end{figure}

\begin{figure}[hbtp]
\includegraphics[scale=0.44]{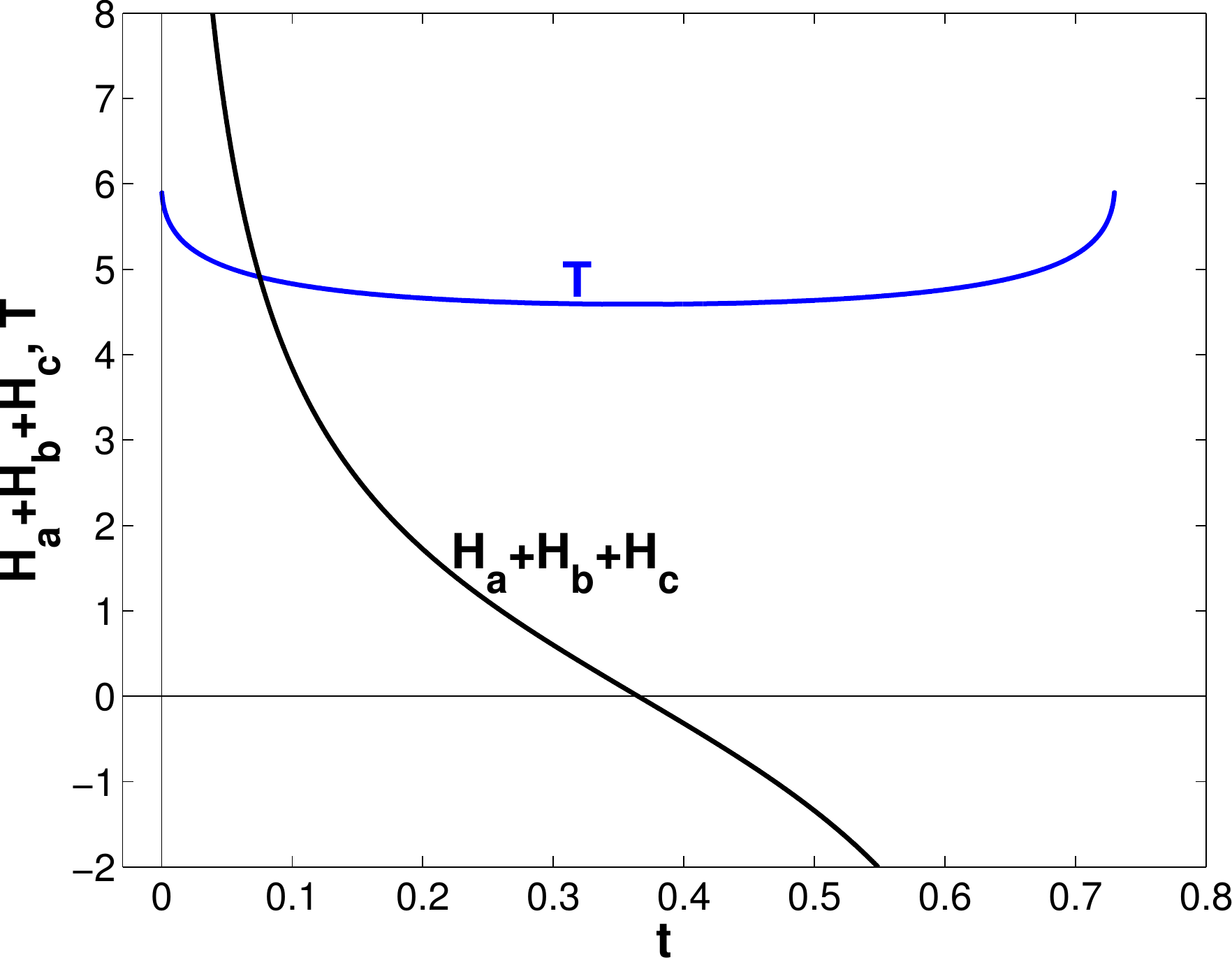}\qquad
\includegraphics[scale=0.44]{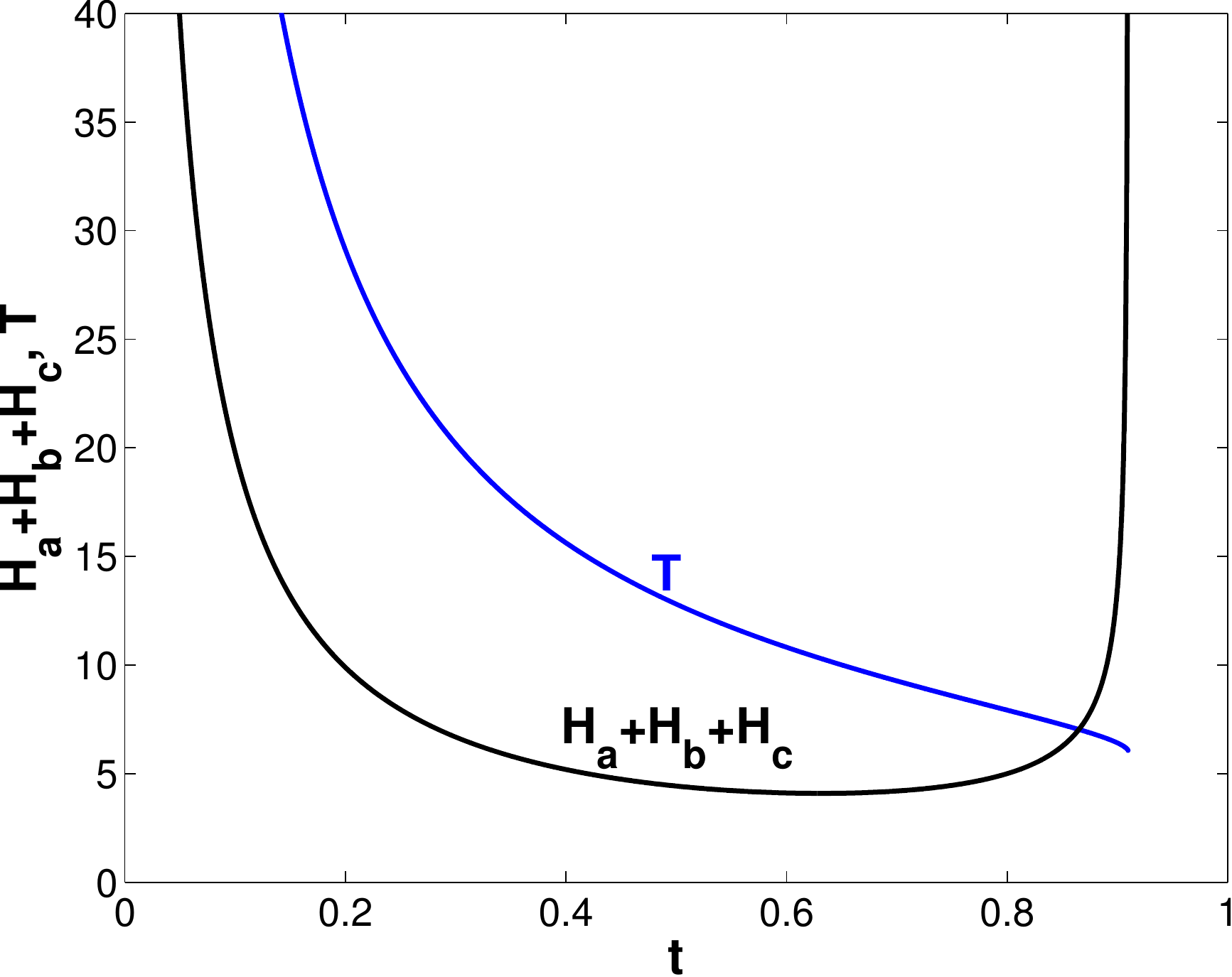} 
\caption{The evolution of the torsion scalar $T$ and sum of Hubble parameters $X=H_a+H_b+H_c$ for $N=2$, $f_0=-\frac{1}{12}$, $w=0$, $K=1$. The initial values are $H_a(0)=157$, $H_b(0)=29.43$, $H_c(0)=-24.8$ for the left graph and $H_a(0)=17630$, $H_b(0)=2670$, $H_c(0)=-2320$ for the right picture. The left plot displays the cosmological scenario \textbf{IIb} and the right one shows the scenario \textbf{IIc}.}
\label{Fig4}
\end{figure}

\begin{figure}[hbtp]
~~~~~~~~~~~~\includegraphics[scale=0.68]{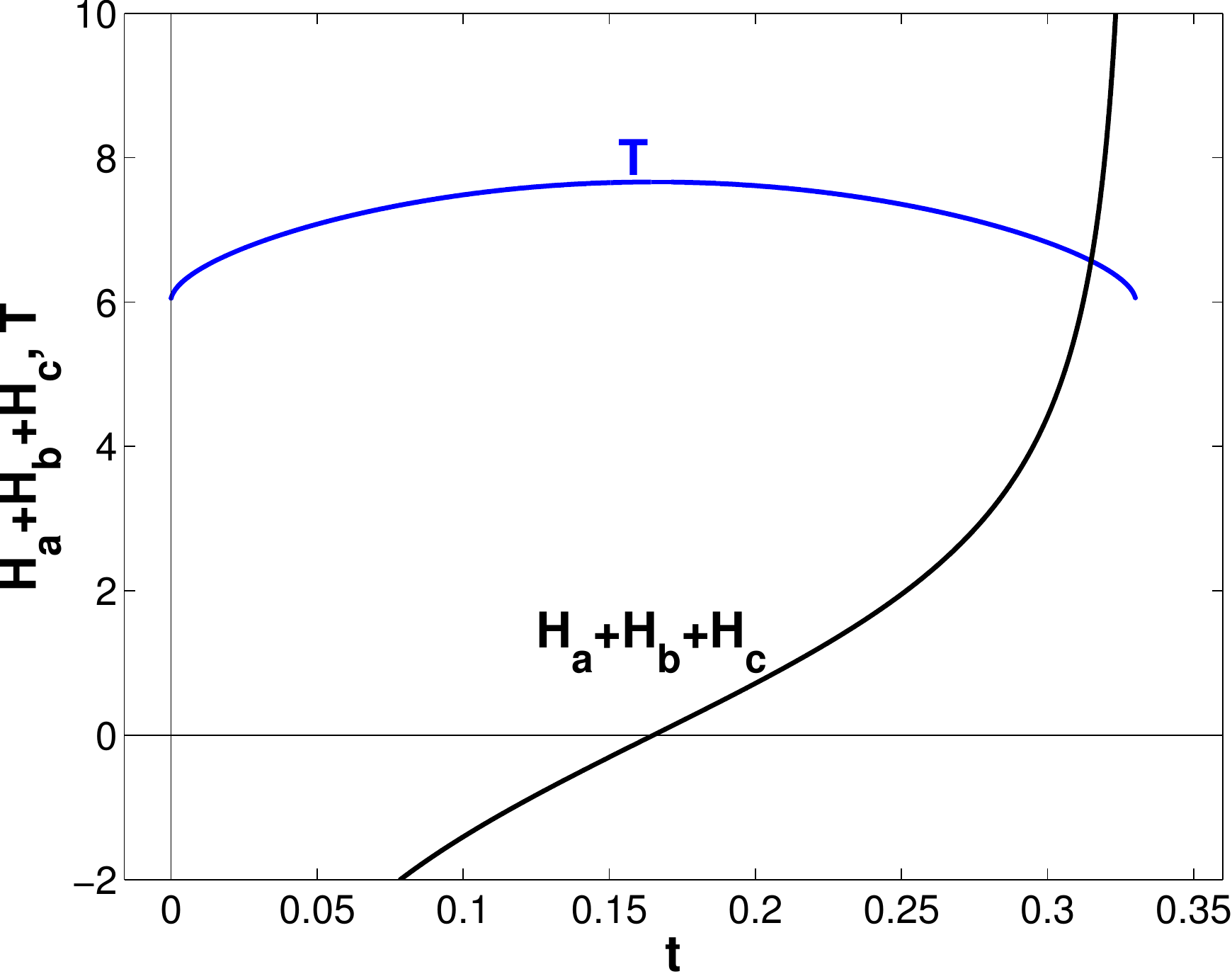}
\caption{The evolution of the torsion scalar $T$ and sum of Hubble parameters $X=H_a+H_b+H_c$ for $N=2$, $f_0=-\frac{1}{12}$, $w=\frac{2}{3}$, $K=1$. The initial values are $H_a(0)=24.7$, $H_b(0)=-44.41$, $H_c(0)=-55.5$. This picture shows the cosmological scenario \textbf{IId}.}
\label{Fig5}
\end{figure}

    In the $f_0<0$ case points \textbf{1}, \textbf{2} and the line \textbf{5} becomes accessible.  We have found, however, one more regime which is not covered by the fixed points listed in the Sect.~3. Namely, this is a non-standard singular regime when $T$ approaches the point where $f_T=\frac{df(T)}{dT}=0$. Our numerical results indicate that for $f_0<0$ four scenarios exist:
\begin{itemize}
\item[\textbf{IIa.}] along the branch $T_1$ from one of three solutions (corresponding to the stationary points \textbf{1}, \textbf{2} or the line \textbf{5} depending on $w$) with infinite $T$ to the isotropic solution of the fixed point~\textbf{3} (see Fig.~\ref{Fig3});
\item[\textbf{IIb.}] along the branch $T_2$ with the start and the finish at the non-standard singularity with $T_{cr}=\frac{1}{-2f_0}$, where $T\to T_{cr}-0$ (see the left picture in Fig.~\ref{Fig4});
\item[\textbf{IIc.}] along the branch $T_2$ from the anisotropic solution of the fixed point \textbf{1} for $w<\frac{1}{2N-1}$ to the non-standard singularity with $T_{cr}=\frac{1}{-2f_0}$, where $T\to T_{cr}+0$ (see the right plot in Fig.~\ref{Fig4});
\item[\textbf{IId.}] along the branch $T_2$ with the start and the finish at the non-standard singularity with $T_{cr}=\frac{1}{-2f_0}$, where $T\to T_{cr}+0$ (see Fig.~\ref{Fig5}).
\end{itemize}

    The scenarios similar to \textbf{Ia}, \textbf{Ib}, \textbf{IIa} are possible for an isotropic Universe and have been described in \cite{Laur}. The scenarios \textbf{IIb}, \textbf{IIc}, \textbf{IId} are intrinsically anisotropic and have no isotropic analogs. It should be noted that the quantity $X=H_a+H_b+H_c=\frac{\dot V}{V}$, where $V\equiv a(t)b(t)c(t)$ is the volume factor. When $X=0$ the cosmological evolution passes through the turnaround (if $\dot X<0$) or the bounce (if $\dot X >0$). The turnaround exists in the case \textbf{IIb} and the bounce occurs in the scenario \textbf{IId}.     
\begin{figure}[hbtp]
\includegraphics[scale=0.44]{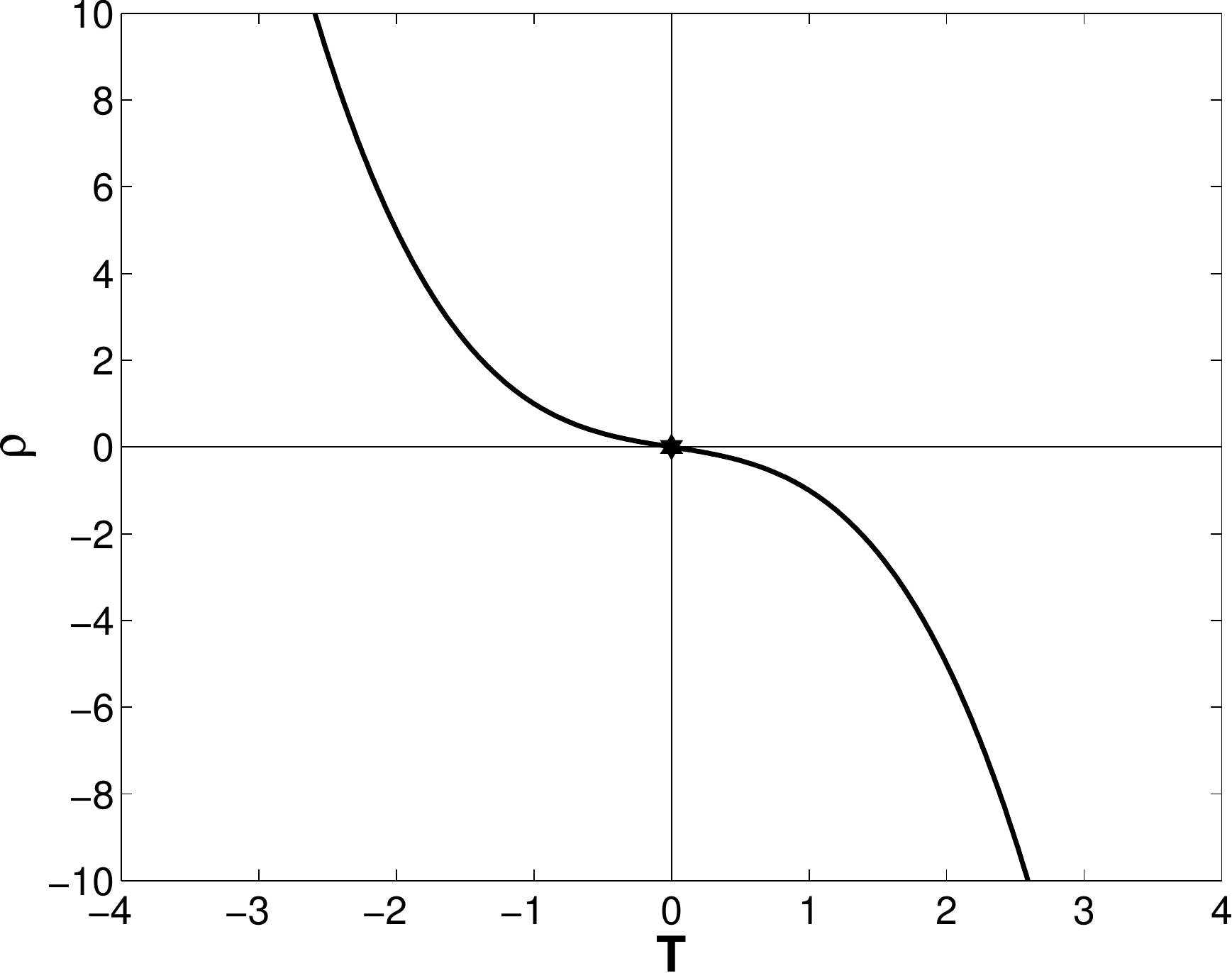}\qquad
\includegraphics[scale=0.44]{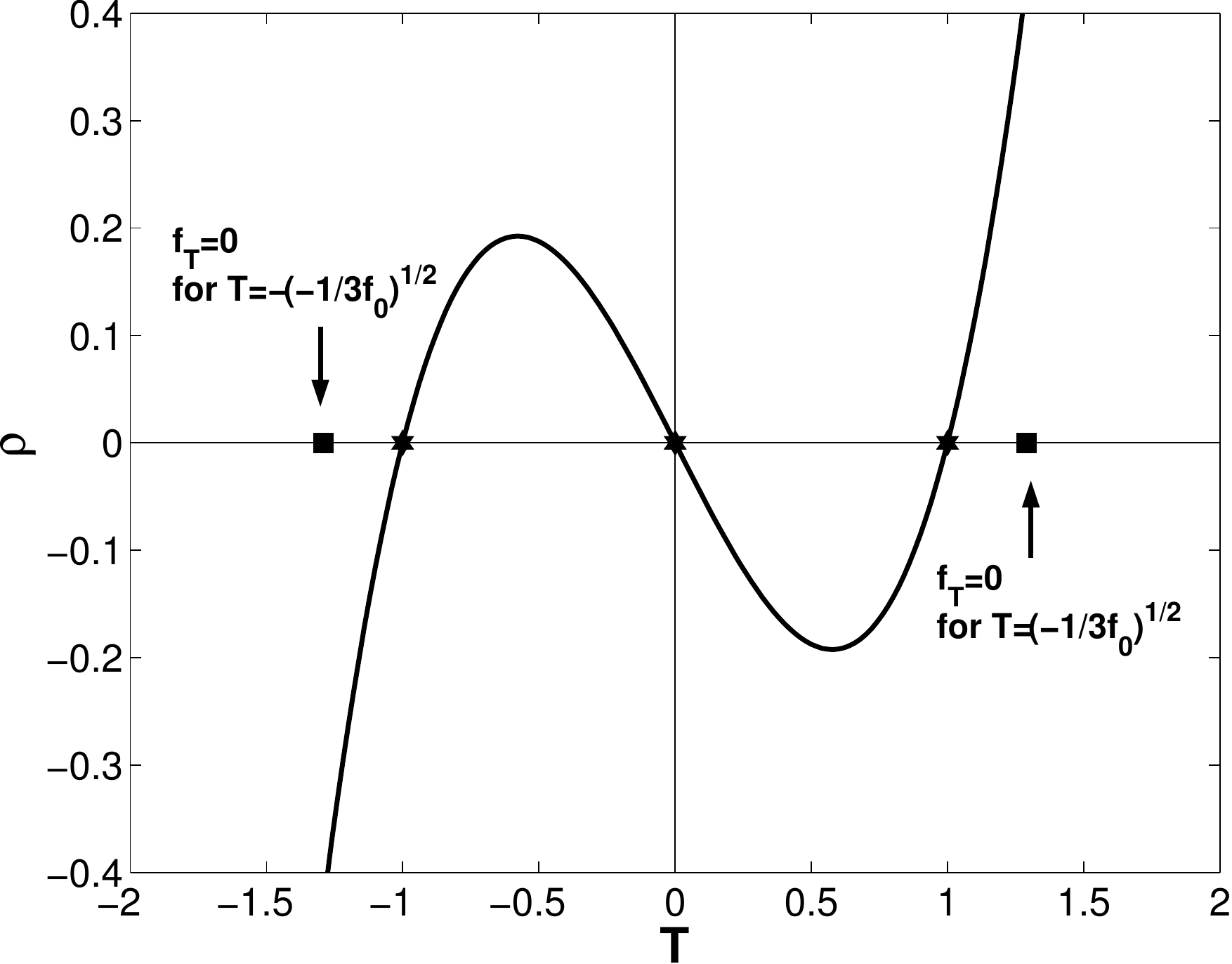} 
\caption{The dependence $\rho(T)=\frac{-T-5f_0T^3}{2K}$, which follows from the constraint equation (\ref{constraint1}) for $N=3$. The parameter $f_0=\frac{1}{5}$ for the left picture, $f_0=-\frac{1}{5}$ for the right plot and for both graphs $K=1$.}
\label{Fig6}
\end{figure}

    For the $N=3$ case the function $\rho(T)=\frac{-T+(1-2N)f_0T^N}{2K}$ is shown in Fig.~\ref{Fig6}. In this case the condition $\rho\geqslant 0$ gives us allowed values of $T$:
\\
\\1). for $f_0>0$ the torsion scalar $T\in[-\infty; 0]$,
\\2). for $f_0<0$ the torsion scalar $T\in\Big{(}-\frac{1}{\sqrt{-5f_0}}; 0\Big{]}\cup\Big{[}\frac{1}{\sqrt{-5f_0}}; +\infty\Big{)}$.
\\
\\From Fig.~\ref{Fig6} we see that only an analog of the scenario \textbf{IIa} is possible for $f_0>0$. On the other hand, analogs of all remaining regimes found above for $N=2$ are possible for $N=3$ with $f_0<0$. We have found all of them in our numerical studies.      
\section{Conclusions}
~~~~We have considered a cosmological evolution of a flat anisotropic Universe in $f(T)=T+f_0 T^N$ gravity, with integer $N>1$ in the presence of a perfect fluid. Combining analytical and numerical methods we have identified possible asymptotic regimes. It is interesting that the Kasner solution does not appear in our list. We know that Kasner solution is an asymptotic regime (in the high-energy limit) for a vacuum cosmological models in $f(T)$ gravity. In GR it is an exact solution for vacuum models and an asymptotic solution for models with perfect fluid (apart from the stiff fluid). This means that the matter content of the Universe in $f(T)$ gravity, in contrast to GR, plays an important role in anisotropic dynamics near a cosmological singularity. For a power-law functions $f(T)=T+f_0T^N$ there exists a critical equation of state parameter for the matter $w_{cr}=1/(2N-1)$ such that for $w>w_{cr}$ the initial cosmological singularity appears to be isotropic. For $w<w_{cr}$ changes in comparison with GR are less drastic --- the initial singularity is still anisotropic, leading terms in the time dependence of Hubble parameters are still power-law type, though the sum of the power indices is in general bigger than $1$. Moreover, for a particular sign of $f_0$ (positive for even $N$ and negative for odd $N$) the nature of the singularity changes completely from the standard one (where Hubble parameters as well as the matter energy density tend to infinity) to the non-standard one (where Hubble parameters and the matter energy density remain finite, and their time derivatives diverge). It is interesting that the de Sitter point exists only in such a situation. As a result, the torsion scalar $T$ can change at most two times during a cosmological evolution ending at de Sitter attractor. All these results indicate also that the famous GR situation described by the phrase ``matter does not matter'', which describes the anisotropic cosmological dynamics near a cosmological singularity, does not occur in $f(T)$ gravity for any equation of state parameter $w$ bigger than $-1$.

\section*{Acknowledgements}
~~~~M. A. S. and A. V. T. are supported by RSF Grant \textnumero 16-12-10401 and A. V. T. is supported by the Russian Government Program of Competitive Growth of Kazan Federal University.

\section{Appendix: The system for numerical integration}
~~~~Using Eq.~(\ref{Hct}) for $\dot H_c$ we rewrite Eq.~(\ref{TtHabct}) for $\dot T$:
\begin{equation}
\label{TtHabct1}
\dot T=-\frac{4}{H_a-H_b}\Big{(}\dot H_a (H_aH_c-{H_b}^2)+\dot H_b({H_a}^2-H_bH_c)\Big{)},
\end{equation}
where $H_a\neq H_b$, and we substitute (\ref{Hct}) and (\ref{TtHabct1}) to Eqs.~(\ref{system11}), (\ref{system21}),
\begin{equation}
\label{Hat}
\frac{4H_af_{TT}}{H_a-H_b}\Big{(}\dot H_a(H_aH_c-{H_b}^2)+\dot H_b({H_a}^2-H_bH_c)\Big{)}+f_T\left( -\dot H_a-{H_a}^2+H_b H_c\right)=\frac{1}{2}K(w+1)\rho, 
\end{equation}   
\begin{equation}
\label{Hbt}
\frac{4H_bf_{TT}}{H_a-H_b}\Big{(}\dot H_a(H_aH_c-{H_b}^2)+\dot H_b({H_a}^2-H_bH_c)\Big{)}+f_T\left( -\dot H_b-{H_b}^2+H_a H_c\right)=\frac{1}{2}K(w+1)\rho. 
\end{equation} 
Collecting terms with $\dot H_a$ and $\dot H_b$ we find
\begin{equation}
\label{Hat1}
\begin{array}{l}
\frac{\dot H_a}{H_a-H_b}\Big{(}4H_af_{TT}(H_aH_c-{H_b}^2)-f_T(H_a-H_b)\Big{)}+\frac{\dot H_b}{H_a-H_b}4H_af_{TT}({H_a}^2-H_bH_c)+\\
+f_T(H_b H_c-{H_a}^2)-\frac{1}{2}K(w+1)\rho=0, 
\end{array}
\end{equation}   
\begin{equation}
\label{Hbt1}
\begin{array}{l}
\frac{\dot H_a}{H_a-H_b}4H_bf_{TT}(H_aH_c-{H_b}^2)+\frac{\dot H_b}{H_a-H_b}\Big{(}4H_bf_{TT}({H_a}^2-H_bH_c)-f_T(H_a-H_b)\Big{)}+\\
+f_T(H_a H_c-{H_b}^2)-\frac{1}{2}K(w+1)\rho=0. 
\end{array}
\end{equation} 
In order to express $\dot H_a$ from this system we multiply Eq.~(\ref{Hat1}) to \\$\Big{(}f_T(H_a-H_b)-4H_bf_{TT}({H_a}^2-H_bH_c)\Big{)}$, Eq.~(\ref{Hbt1}) by $4H_af_{TT}({H_a}^2-H_bH_c)$ and then we sum the obtained results,
\begin{equation}
\label{Hat2}
\begin{array}{l}
\frac{\dot H_a}{H_a-H_b}\Big{(}4H_af_{TT}(H_aH_c-{H_b}^2)-f_T(H_a-H_b)\Big{)}\Big{(}f_T(H_a-H_b)-4H_bf_{TT}({H_a}^2-H_bH_c)\Big{)}+\\
+\frac{\dot H_a}{H_a-H_b}16H_aH_b{f_{TT}}^2(H_aH_c-{H_b}^2)({H_a}^2-H_bH_c)+\\
+\Big{(}f_T(H_b H_c-{H_a}^2)-\frac{1}{2}K(w+1)\rho\Big{)}\Big{(}f_T(H_a-H_b)-4H_bf_{TT}({H_a}^2-H_bH_c)\Big{)}+\\
+4H_af_{TT}({H_a}^2-H_bH_c)\Big{(}f_T(H_a H_c-{H_b}^2)-\frac{1}{2}K(w+1)\rho\Big{)}=0.
\end{array}
\end{equation}
It is transformed to   
\begin{equation}
\label{Hat3}
\begin{array}{l}
\dot H_af_T\Big{[}4f_{TT}\Big{(}H_a(H_aH_c-{H_b}^2)+H_b({H_a}^2-H_bH_c)\Big{)}-f_T(H_a-H_b)\Big{]}=\\
=-\Big{(}f_T(H_b H_c-{H_a}^2)-\frac{1}{2}K(w+1)\rho\Big{)}\Big{(}f_T(H_a-H_b)-4H_bf_{TT}({H_a}^2-H_bH_c)\Big{)}-\\
-4H_af_{TT}({H_a}^2-H_bH_c)\Big{(}f_T(H_a H_c-{H_b}^2)-\frac{1}{2}K(w+1)\rho\Big{)}. 
\end{array}
\end{equation} 
Finally, we have
\begin{equation}
\label{Hat4}
\begin{array}{l}
\dot H_a=\frac{\Big{(}H_b H_c-{H_a}^2-\frac{1}{2f_T}K(w+1)\rho\Big{)}\Big{(}f_T(H_a-H_b)-4H_bf_{TT}({H_a}^2-H_bH_c)\Big{)}}{f_T(H_a-H_b)-4f_{TT}\Big{(}{H_a}^2(H_b+H_c)-{H_b}^2(H_a+H_c)\Big{)}}+\\
+\frac{4H_af_{TT}({H_a}^2-H_bH_c)\Big{(}H_a H_c-{H_b}^2-\frac{1}{2f_T}K(w+1)\rho\Big{)}}{f_T(H_a-H_b)-4f_{TT}\Big{(}{H_a}^2(H_b+H_c)-{H_b}^2(H_a+H_c)\Big{)}}. 
\end{array}
\end{equation} 
The expression for $\dot H_b$ is derived similarly to $\dot H_a$. It has the form
\begin{equation}
\label{Hbt2}
\begin{array}{l}
\dot H_b=\frac{4H_bf_{TT}(H_aH_c-{H_b}^2)\Big{(}H_bH_c-{H_a}^2-\frac{1}{2f_T}K(w+1)\rho\Big{)}}{f_T(H_a-H_b)-4f_{TT}\Big{(}{H_a}^2(H_b+H_c)-{H_b}^2(H_a+H_c)\Big{)}}+\\
+\frac{\Big{(}H_a H_c-{H_b}^2-\frac{1}{2f_T}K(w+1)\rho\Big{)}\Big{(}f_T(H_a-H_b)-4H_af_{TT}(H_aH_c-{H_b}^2)\Big{)}}{f_T(H_a-H_b)-4f_{TT}\Big{(}{H_a}^2(H_b+H_c)-{H_b}^2(H_a+H_c)\Big{)}}. 
\end{array}
\end{equation} 
Substituting Eqs.~(\ref{Hat4}), (\ref{Hbt2}) to Eq.~(\ref{Hct}) we find
\begin{equation}
\label{Hct1}
\begin{array}{l}
\dot H_c=\frac{(H_c-H_b)\Big{(}H_b H_c-{H_a}^2-\frac{1}{2f_T}K(w+1)\rho\Big{)}\Big{(}f_T(H_a-H_b)-4H_bf_{TT}({H_a}^2-H_bH_c)\Big{)}}{(H_a-H_b)\Big{[}f_T(H_a-H_b)-4f_{TT}\Big{(}{H_a}^2(H_b+H_c)-{H_b}^2(H_a+H_c)\Big{)}\Big{]}}+\\
+\frac{4H_af_{TT}(H_c-H_b)({H_a}^2-H_bH_c)\Big{(}H_a H_c-{H_b}^2-\frac{1}{2f_T}K(w+1)\rho\Big{)}}{(H_a-H_b)\Big{[}f_T(H_a-H_b)-4f_{TT}\Big{(}{H_a}^2(H_b+H_c)-{H_b}^2(H_a+H_c)\Big{)}\Big{]}}+\\
+\frac{4H_bf_{TT}(H_a-H_c)(H_aH_c-{H_b}^2)\Big{(}H_bH_c-{H_a}^2-\frac{1}{2f_T}K(w+1)\rho\Big{)}}{(H_a-H_b)\Big{[}f_T(H_a-H_b)-4f_{TT}\Big{(}{H_a}^2(H_b+H_c)-{H_b}^2(H_a+H_c)\Big{)}\Big{]}}+\\
+\frac{(H_a-H_c)\Big{(}H_a H_c-{H_b}^2-\frac{1}{2f_T}K(w+1)\rho\Big{)}\Big{(}f_T(H_a-H_b)-4H_af_{TT}(H_aH_c-{H_b}^2)\Big{)}}{(H_a-H_b)\Big{[}f_T(H_a-H_b)-4f_{TT}\Big{(}{H_a}^2(H_b+H_c)-{H_b}^2(H_a+H_c)\Big{)}\Big{]}}.
\end{array}
\end{equation}
The system of equations (\ref{Hat4})-(\ref{Hct1}), is used for the numerical integration, where $\rho=\frac{1}{2K}(f(T)-2Tf_T)$, the function $f(T)$, its derivatives $f_T$, $f_{TT}$ and $T=-2(H_aH_b+H_aH_c+H_bH_c)$ should be substituted.

\end{document}